# Ion implantation in nanodiamonds: size effect and energy dependence

Andrey A. Shiryaev[1,2*], Jonathan A. Hinks[3], Nigel A. Marks[4], Graeme Greaves[3], Felipe J. Valencia[5,6,7], Stephen E. Donnelly[3], Rafael I. González[7,8], Miguel Kiwi[6,7], Alexander L. Trigub[9], Eduardo M. Bringa[10,11], Jason L. Fogg[4], Igor I. Vlasov[12,13]

1) Frumkin Institute of Physical Chemistry and Electrochemistry RAS, Leninsky pr .31 korp. 4, Moscow 119071, Russia. Tel. +7 495 9554664.
2) Chemistry Dept., Lomonosov Moscow State University, Moscow, Russia
3) University of Huddersfield, Queensgate, Huddersfield, HD1 3DH, United Kingdom
4) Dept. of Physics and Astronomy, Curtin University, Perth, Australia
5) Núcleo de Matemáticas, Física y Estadística, Facultad de Ciencias, Universidad Mayor, Chile
6) Departamento de Física, Facultad de Ciencias, Universidad de Chile, Casilla 653, Santiago, Chile
7) Centro para el Desarrollo de la Nanociencia y la Nanotecnología, CEDENNA, Avda. Ecuador 3493, Santiago, Chile 9170124
8) Centro de Nanotecnología Aplicada, Facultad de Ciencias, Universidad Mayor, Camino La Pirámide 5750, Huechuraba, Santiago
9) National Research Center Kurchatov Institute, Moscow, Russia
10) Facultad de Ciencias Exactas y Naturales, Universidad Nacional de Cuyo, Mendoza 5500, Argentina
11) CONICET, Argentina
12) General Physics Institute RAS, Vavilova St. 38, Moscow, Russia
13) National Research Nuclear University MEPhI, Moscow, 115409, Russia

*Corresponding author*: Email: shiryaev@phyche.ac.ru; a_shiryaev@mail.ru (Andrey Shiryaev)

**Abstract**

Nanoparticles are ubiquitous in nature and are increasingly important for technology. They are subject to bombardment by ionizing radiation in a diverse range of environments. In particular, nanodiamonds represent a variety of nanoparticles of significant fundamental and applied interest. Here we present a combined experimental and computational study of the behaviour of nanodiamonds under irradiation by xenon ions. Unexpectedly, we observed a pronounced size effect on the radiation resistance of the nanodiamonds: particles larger than 8 nm behave similarly to macroscopic diamond (i.e. characterized by high radiation resistance) whereas smaller particles can be completely destroyed by a single impact from an ion in a defined energy range. This latter observation is explained by extreme heating of the nanodiamonds by the penetrating ion. The obtained results are not limited to nanodiamonds, making them of interest for several fields, putting constraints on processes for the controlled modification of nanodiamonds, on the survival of dust in astrophysical environments, and on the behaviour of actinides released from nuclear waste into the environment.

*1. Introduction*

The interaction of energetic particles with nanomaterials is of considerable interest for fields ranging from nanotechnology to nuclear materials [1] to astrophysics [2]. Fundamental questions about the size-dependence of radiation resistance remain unanswered; whereas some studies indicate higher resistance with decreasing size, others show the inverse behaviour [3]. Irradiation of, and implantation into, nanosized diamonds is of importance for the controlled formation of luminescent nanoparticles for biomedicine and quantum computing [4, 5]. Besides technological applications, ion implantation is the most plausible mechanism to explain the cosmochemistry of trace elements in nanodiamonds (NDs) extracted from meteorites [6]. Bulk diamond is characterized by high radiation resistance [7] but relatively little is known about the behaviour of NDs under irradiation. The effects of ionizing radiation [8] and of swift heavy ions [9, 10] on ND films have been reported.

Among the elements which can be introduced into a diamond lattice, xenon is of particular importance for solid-state physics as well as for cosmochemistry and astrophysics. Xenon impurities can couple with a vacancy in the diamond lattice to form a stable defect (Xe-V) characterized by a narrow zero-phonon line in the near-infra-red luminescence spectra [11] which is a potential candidate as a source of single photons and optically manipulated qubits. Studies of Xe isotopes in meteoritic NDs provide evidence for a strong contribution from supernovae ejecta thus indicating that at least some meteoritic NDs are presolar [12]. The statistical distributions of implanted Kr and Xe ions in NDs after ballistic stage have been modelled [13,14] using a Monte-Carlo SRIM code [15]. Complicated bimodal patterns of xenon release upon heating from meteoritic and ion-implanted synthetic NDs indicate the existence of at least two structural sites [16].

Here, we present results of an *in situ* Transmission Electron Microscopy (TEM) investigation of implantation of low energy (6 and 40 keV) Xe ions into dispersed nanodiamonds spanning a size range between ~2 and 40 nm. This is complemented by molecular dynamics simulations of the implantation process and by quantum chemistry calculations of the stability of Xe-related defects. It is shown that at sizes below 8 nm, a ND grain may be completely destroyed by the impact of a small of ions or even by a single impact provided that the ion energy is around 6 keV. The magnitude of the effect strongly depends on the nature of the impact, i.e. central collisions destroy the grain, whereas the glancing ones incur much smaller damage. Larger NDs behave like bulk diamond by demonstrating a high radiation resistance. These results put constraints both on the conditions for the implantation processes intended for the controlled modification of NDs and on the astrophysical environments where the implantation of noble gas ions may have taken place. This work shows the importance of the heating of small NDs by impinging ions and contributes to the understanding of the survival of cosmic dust.

*Results*

*In situ Xe ions implantation in TEM*

Figures 1 and 2 show TEM micrographs of detonation and meteoritic nanodiamonds (see Methods section) before and after irradiation with 6 keV Xe at room temperature to a fluence of ~$6\times10^{14}$ ions/cm$^2$. All the images were recorded during *in situ* implantation in a TEM at the MIAMI facility at the University of Huddersfield [17]. Several differences such as disappearance of small grains (Figure 1 a,c); a decrease of contrast suggesting mass loss and decrease of crystallinity (Figure 1 b, d) are immediately observable. Figure 2 shows analysis of the evolution of nanodiamond grains of different sizes obtained by pixel-by-pixel analysis of individual grains which are clearly free from overlaps. Remarkably, whereas rarer large (>10 nm) grains survive the experiment, the smaller grains gradually disappear.

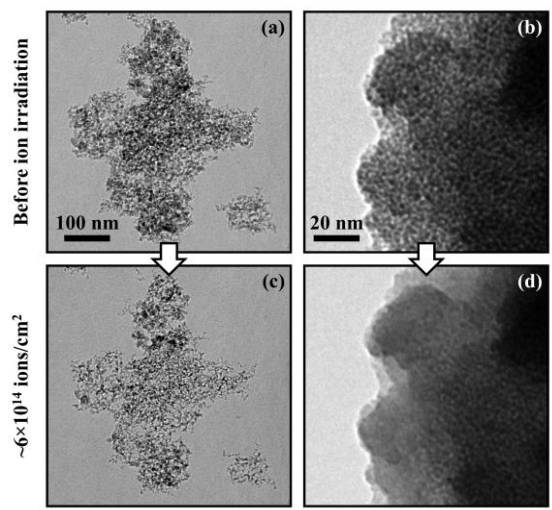

**Figure 1.** TEM images of detonation (a+c) and meteoritic (b+d) NDs before (a+b) and after (c+d) irradiation with 6 keV Xe to ~$6\times10^{14}$ ions/cm$^2$ at room temperature in the MIAMI facility. The detonation NDs shown had an average diameter of 5 nm whereas the meteoritic NDs shown were smaller with an average of 2 nm. Examples of the formation of carbon "ribbons" during the ion irradiation can be seen (c).

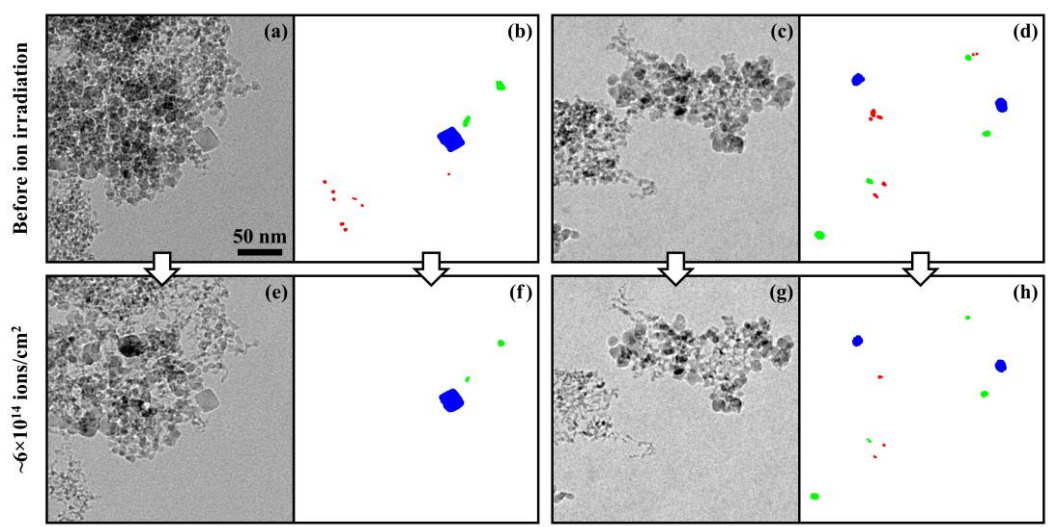

**Figure 2.** TEM images and analysis of detonation NDs on a carbon-film support before (a–d) and after (e–h) irradiation with 6 keV Xe to ~6×10$^{14}$ ions/cm$^2$ at room temperature in the MIAMI facility. Individually-identifiable NDs on the peripheries of the clusters are highlighted according to their initial diameters in red (<5 nm), green (~10 nm) and blue (>15 nm). As can be seen, the majority of the smallest NDs which could be tracked were found to disappear completely and the sputtering effects were observed to become less-pronounced for larger particle sizes. Examples of the formation of carbon "ribbons" during the ion irradiation can be seen in both (e) and (g). The scale marker in (a) applies to all the panels in the figure.

Azimuthally integrated electron diffraction patterns as well as EELS spectra acquired before and after the implantation show only minor changes which are difficult to quantify due to changes in sample thickness. However, whereas the diffraction pattern of the unirradiated sample featured uniform rings (i.e. pure powder diffraction), after the irradiation diffraction spots from individual grains superposed on the rings became more distinct. This behaviour can be explained by the decreased number of ND grains in the beam. Note that these changes observed in the diffraction patterns became obvious only after relatively large ion fluences (~10$^{16}$ ions/cm$^2$).

The implantation effects on clumps of meteoritic diamonds possessing the smallest average sizes are mostly manifested as a loss of contrast and individual grains were not resolved above a 6 keV ion fluence of ~5.6×10$^{14}$ ions/cm$^{-2}$ (Fig. 1 b,d). This effect was observed also at 40 keV.

In contrast to the two aforementioned cases, the irradiation of the 40 nm NDs led to only minor changes which can be described as the destruction of poorly ordered carbon on the grain surfaces. The size of the grains remained constant and all the monitored grains survived the irradiation up to the maximal fluences studied (a few ×10$^{16}$ ions/cm$^2$).

Whilst our TEM observations of the irradiation response of the larger NDs can be attributed to conventional sputtering processes, an explanation for the behaviour of the smallest NDs requires alternative mechanisms to be considered. In order to explore the underlying processes responsible, we have complemented our experimental data by state of the art computational modelling as discussed below.

*Molecular dynamics modelling*

Molecular Dynamics (MD) modelling was performed in order to evaluate the effects of heavy ion irradiation using two independent approaches for the C-C interaction (see Methods section). The principal result of both calculation strategies is a significant increase in temperature due to the energy transferred to the nanoparticle during the ballistic phase of the Xe implantation. As shown in Fig. 3, NDs up to 4 nm are heated to extremely high temperatures (above 3000 K, see also Fig. S1). Due to poor thermal contact with surrounding grains and low thermal emissivity of dielectric nanoparticles [18], the impacted nanodiamond remains hot for several picoseconds (Fig. S2), and consequently undergoes self-annealing. It is tempting to term this self-annealing process a thermal spike, but even though the NDs get very hot, the heating process is distinct from conventional thermal spike behaviour well-known in metals and oxides. In those systems localized melting occurs in a fraction of a picosecond and is induced by the cascade evolution, whereas here the heating is driven by the finite size of the nanoparticle itself, as there is no heat loss path to an external reservoir. Indeed, recent EDIP-MD simulations of collision cascades in bulk diamond show that cascades produce fractal-like trajectories and point defects, without the slightest hint of melting [19]. Accordingly, the effect seen here can be definitely attributed to the nanoparticle itself and should be understood as a radiation effect specific to small objects.

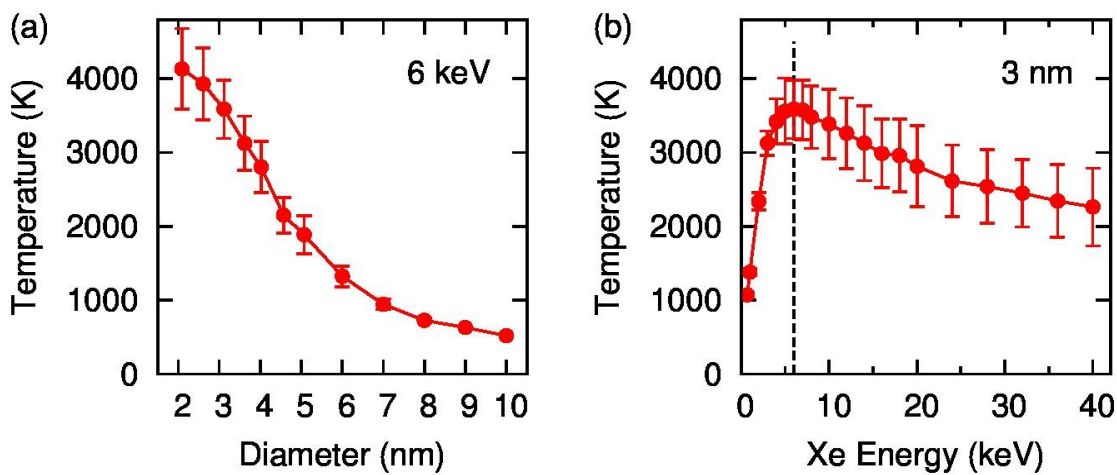

**Figure 3.** Peak temperature of ND grains as functions of diameter after a 6 keV Xe ion impact (a) and of incident Xe energy in a 3 nm ND (b).

The confinement of kinetic energy within the ND is quantified in Fig. 4 which shows the amount of energy ejected from the ND. Figure 4a shows that for a small 2 nm particle most of the initial 6 keV of energy is not retained by the ND, with the Xe typically exiting the ND with an energy ≥2 keV, and another ≥2 keV removed by high-velocity carbon atoms. However, for a 4.5 nm ND the total ejected energy is about 1 keV leaving 5 keV of the kinetic energy of the Xe ion free to be distributed into the ND, where it primarily contributes to atomic motion (i.e. temperature). This ability of large NDs to capture all of the Xe energy is the reason why the temperatures in Fig. 3a do not fall off proportionally to the number of atoms (i.e as the third power of diameter), but instead show a much more gradual reduction. The efficiency of energy transfer is a strong function of the Xe energy as seen in Figs. 3b and 4b. The dotted line shows that maximum heating occurs for 6 keV, and falls off significantly as the Primary knock-on atoms (PKA) energy increases. This reduction in transfer efficiency is reflected in Fig. 4b where the energy of the exiting Xe atoms increases linearly, with no additional residual energy transfer to the ND.

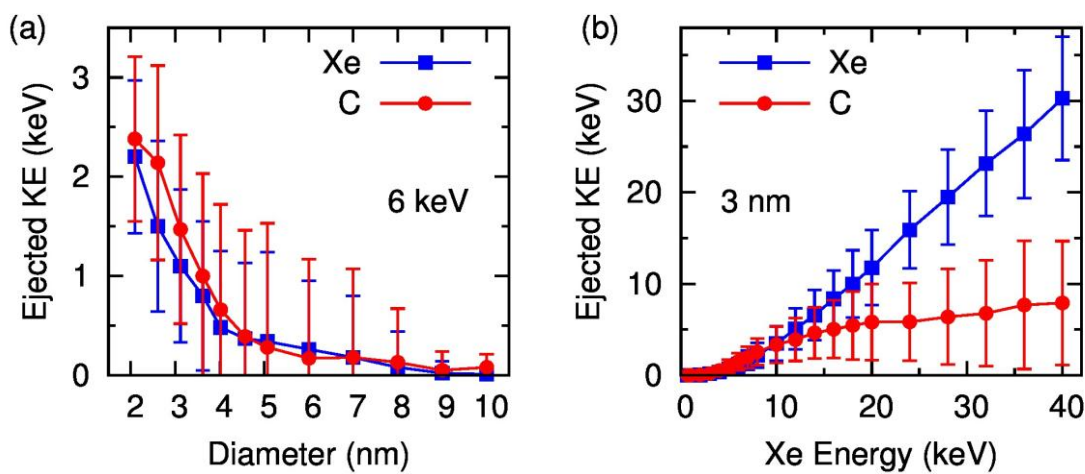

**Figure 4.** Kinetic energy of ejected Xe and C atoms as functions of diameter for a 6 keV ion (a) and of incident Xe energy for a 3 nm ND (b).

Figure 5 shows the effect of the nanoparticle heating on the diamond-like character of the cluster. Since the temperatures are approaching the melting point of diamond (~4300 K) [20], the particles undergo extensive structural reorganization. Figure 6 shows the $sp^3$ fraction in the original state and several picoseconds after the impact. For small particles around 2 nm a single impact is sufficient to remove the majority of the $sp^3$ bonding, destroying the diamond structure (see Supplementary movies). With increasing particle size, the loss of diamond character is reduced, requiring multiple Xe impacts to severely disrupt the ND, while for particles above 8 nm the sheer number of atoms in the particle means the ND is highly resistant to any heating-induced damage.

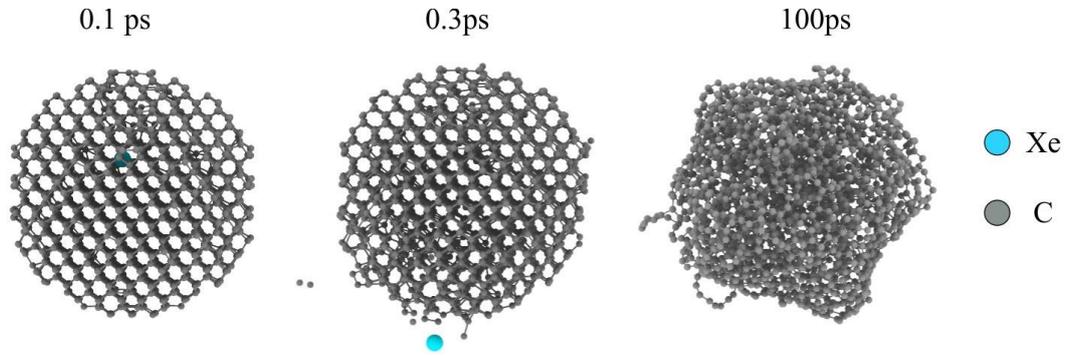

**Figure 5.** Visualizations of MD modelling results of a 3 nm ND grain at 0.1, 0.3 and 100 ps after a 4 keV Xe ion impact.

Comparison of simulated damage occurring at different energies of Xe ions shows that the heating effect is the strongest at 6 keV. At higher incident energies the total energy of ejected atoms is similar to that of the incident ion. This correlates with the reduction in temperature seen in Fig. 3, where the cluster temperature falls off at high Xe energies.

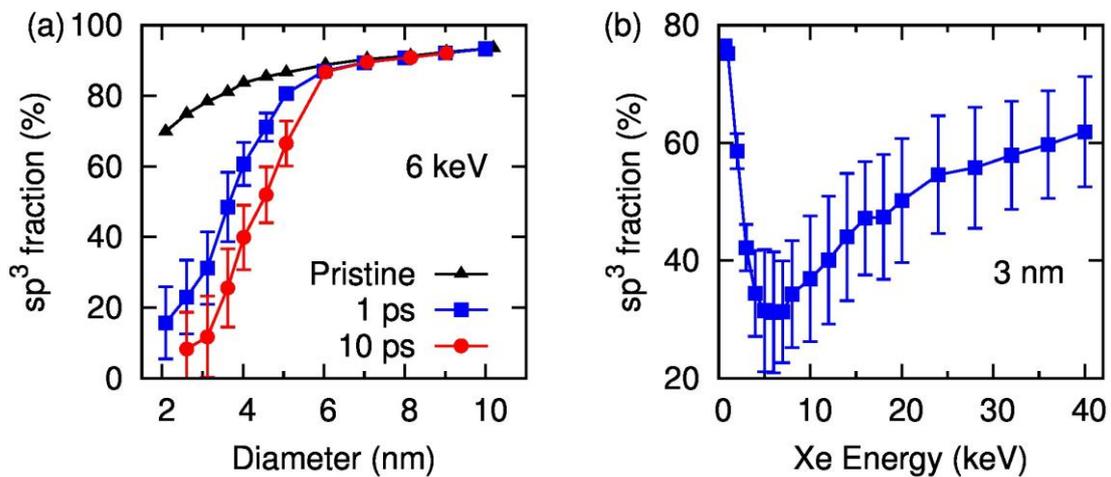

**Figure 6.** Fraction of $sp^3$ bonding as functions of diameter after a 6 keV Xe impact (a) and of incident Xe energy for a 3 nm ND (b).

Modelling of off-axis impact was also performed and Fig. 7 shows the role played by the displacement of the impact spot from centre of a grain (impact parameter **b**) on 5 nm NDs. Figure 7a shows that the temperature increase is strongly dependent on **b**. For the 4 keV Xe ion the **b** = 0 and **b** = 1.25 values are the quite similar, since the ion delivers the energy during the first collision stages (i.e. in the first nanometres). However, for energies larger than 6 keV a gap between the curves opens since for **b** = 1.25 nm the Xe atom crosses the ND and emerges on the opposite side. Figure 7b illustrates the fact that the number of atoms ejected strongly depends on **b** possibly due to the lateral straggling of the cascade. The damage created is the largest in central collisions. In addition, Fig. 7c shows that the sp$^2$/sp$^3$ diminishes as a function of **b**, since the Xe ion interacts with fewer C atoms.

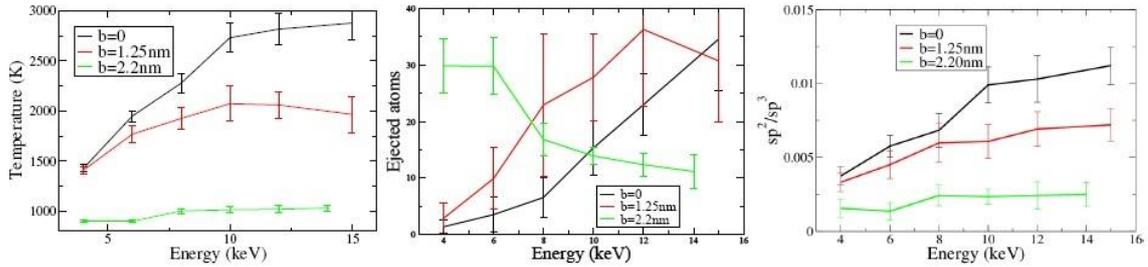

**Figure 7.** Influence of the position of the initial ion impact on the resulting modification of 5 nm ND grains. The offset, *b*, is the distance from the grain centre to the initial impact. See text for details.

*Stability of Xe-Vacancy defects in nanodiamond*

Implantation of Xe with energies below ~1 keV leads to moderate heating, limited modification of the ND and implantation of the Xe ion. The behaviour of implanted Xe was investigated in H-terminated non-spherical nanodiamond particles using the Quickstep module of the CP2K program suite. Calculated values of the Xe-V formation energy $E_{form}$ in the centre of ND were 15.5, 15.4 and 15.8 eV for particles with diameters of 1.6, 2.0 and 3.0 nm, respectively. Therefore, for a wide range of ND sizes the formation energy of a Xe-V defect is independent of the grain size. The stability of the Xe-V defect in a 2 nm particle was estimated by optimization of the defect geometry in different lattice nodes in the {110}, {111} and {001} directions from the centre of the particle towards the surface. In all cases we observed a moderate (1 to 2 eV) decrease of the formation energy for the Xe atoms placed near the surface as shown in Fig. 8. This result indicates that Xe atoms in NDs will tend to diffuse towards the grain surface or to extended defects such as grain boundaries. In diamond this behaviour is demonstrated by N and B impurities [21], but Si impurities behave very differently [22].

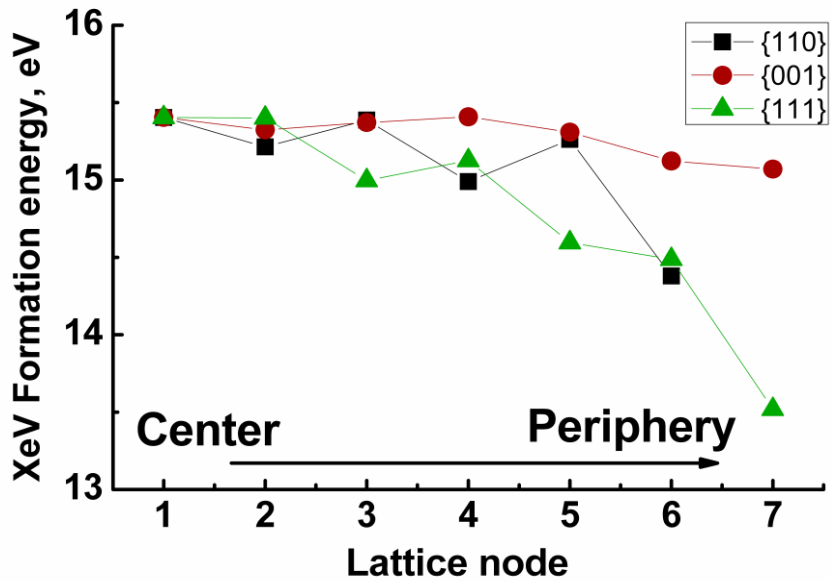

**Figure 8.** Formation energies of a Xe-vacancy defect in a 2 nm ND grain at successive lattice nodes along the [110], [001] and [111] crystallographic directions.

*Discussion*

The results of the experiment and simulations can be summarized as follows: 1) a single 6 keV Xe is sufficient to destroy a ND with dimensions in the range 2 to 3 nm; 2) several Xe impacts are required to destroy a 5 nm ND, but the effect is markedly influenced by the character of the impact (i.e. resulting in a collision cascade in the centre or a much less disrupting glancing collision); and 3) NDs of order 8 nm and higher are characterized by radiation resistance similar to bulk diamond. The observed effect markedly exceeds published results on ion-induced damage in extremely thin (1-3 mg/cm$^2$) flat diamond-like carbon (DLC) foils [23, 24], where only up to 3.5 atoms were removed by a 40 keV Ar$^+$ ions at normal incidence. We believe that the loss of material under ion irradiation observed by us is explained by a particular kind of a sputtering effect, dramatically enhanced by small size of the nanoparticles involved.

The preservation of some 5 nm ND grains during the irradiation experiments even at fluences at which, on average, several ions will have impacted every nm$^2$ can be explained by shadowing from the ion beam by other NDs and our modelling results for off-centre impacts. Figure 7 shows that at grain sizes of 4–6 nm, the temperature reached by an impacted nanoparticle is a strong function of the impact position. Glancing impacts lead to sputtering, but the ND particle may well survive. In addition, ND particles hanging in vacuum on edges of the clumps necessarily possess good thermal contact with the rest of the clump which facilitates cooling. In contrast, loose particles forming large heaps may, on average, have much poorer contact with their surroundings and are thus prone to impact-induced evaporation.

Therefore, the MD simulations may help to explain the origin of the unexpected TEM observation of the destruction of NDs by highlighting the very substantial heating effect caused by the Xe impact. This effect is not captured in the simple Monte-Carlo SRIM code which includes binary collisions only and neglects the heating effect. However, using the number of Frenkel defects calculated with SRIM as a measure of the damage, we estimate that only ions with a mass equal to or greater than that of Kr (84 amu) can lead to comparable radiation effects (particle disruption) whereas under no conditions will lighter ions such as Si or N destroy a ND. The latter conclusion is supported by our Tersoff LAMMPS MD simulations of Si implantation in 5 to 7 nm NDs which show that the ion-induced temperature rise is at least two times smaller than in the Xe case and is insufficient to destroy the ND. However, the sputtering yield does not change significantly. Experiments on N implantation into 12 nm nanodiamond particles support these results [5].

The observed destructive effects are not limited to Xe ions in diamond and that most nanosized materials will react in a qualitatively similar way to irradiation with other heavy ions. For example, dramatic increases in the sputtering yield due to heavy ion irradiation of gold nanorods compared to that from flat foils has been observed and successfully explained by the proximity of collision cascades to the particle surface [25].

*Implications for survival of dust in astrophysical environment*

Survival of dust in space is limited by several factors. In many astrophysical environments sputtering by ionizing radiation plays an important role [26]. Although it is generally assumed that stochastic heating by ultraviolet (UV) photons may strongly heat the smallest grains ($\leq 2$ nm) [2], the influence of radiation on very small grains remains poorly understood. Our results indicate that heating effects due to ion impacts may give rise to qualitatively similar heating effects. Size dependence of the Si-V luminescence intensity of meteoritic NDs may be explained by such processes, as substantial heating of small NDs by Si ion implantation can promote the formation of an optically-active SiV defect [27]. In contrast to the UV-related heating which is primarily determined by the absorption coefficient and grain size, the effect of ions demonstrates remarkable energy dependence with the existence of a "sweet spot". Whereas high-energy ions can pass through a nanoparticle without noticeable effect, ions of a few keV can completely destroy small nanoparticles and heavily damage medium-sized nanoparticles converting them (fully or partially) to $sp^2$-bonded carbons – effects unattainable in the case of UV photons.

As demonstrated by the quantum chemistry modelling, the diamond lattice tends to expel most Xe impurities. This observation is in line with experimental observations that only ~10% of implanted Xe ions are trapped in nanodiamond grains [16]. On heating, NDs release heavy noble gases (Xe, Kr, Ar) in several temperature steps ("components") with somewhat different isotopic compositions. Concentration of the high temperature component (Xe-HL) increases with increasing ND grain size [28]. In view of our calculations one may assign the high temperature components (Xe-HL and Xe-P6) to ions residing in diamond lattice after the implantation, whereas the low temperature release is due to ions outdiffused to grain surfaces or extended defects such as a grain boundary or twin. This scenario explains how the bimodal pattern of noble gases release from presolar diamonds may result from a single implanted component. In the same time, light gases (He, Ne) are largely released in a single broad peak [28]. Impact of light ions (such as He, Ne, N, Si) produce only a moderate temperature increase and mostly result in the creation of point defects, thus a single release peak is observed. Discussion of the isotopic composition of the noble gases [6,12-14,29-30] is beyond the scope of the current paper.

*Conclusions*

Defects in diamonds involving heavy elements, such as Xe and Eu, attract considerable interest due to promising applications in modern nanotechnology [31, 32]. Whereas at present luminescent nanodiamonds are produced mostly from pre-irradiated macro- and microdiamonds, efforts to produce desired defects in situ in nanoparticles already give promising results [4,5]. Our results show that attempts to introduce desired elements into nanoparticles by ion implantation may lead to destruction of grains of certain sizes during implantation; therefore, irradiation conditions should be carefully calculated beforehand.

The effects described in this paper are not limited to nanodiamonds and have a more general character. For example, recoil uranium atoms may disrupt $PuO_2$ colloids and other nanoparticles which play an important role in the environmental behaviour of actinides [33]. The exact outcome will depend on the radiation resistance and thermodynamic stability of a given nanoparticulate phase.

**Methods**

*Samples and in situ TEM*

Nanodiamonds (NDs) with widely different sizes were studied: a) natural NDs extracted from the Orgueil meteorite following standard protocol [12] are characterized by a log-normal size distribution between 1 and 10 nm with median of approximately 2.6 nm [6,12]; b) synthetic NDs obtained by detonation (DNDs) with a rather narrow size distribution peaked between 4 to 5 nm, but with a small number of considerably larger grains [34]; and c) synthetic NDs with grain sizes of 30 to 40 nm obtained from explosives [35]. The nanodiamonds were dispersed on holey-carbon TEM grids from an ethanol solution.

Monte-Carlo calculations [15] of $^{130}$Xe ions impacting on a 5 nm nanodiamond (displacement energy, $E_d$, of 50 eV [36]) show that, at ion energies above 6 keV, the ions penetrate through the grain creating several tens of rather homogeneously distributed Frenkel defects. Based on these calculations we have selected energies of incident energies for *in situ* implantation in TEM performed at the MIAMI facility [17] using 6 and 40 keV Xe ions at room temperature with an electron beam energy of either 80 or 200 keV. The *in situ* nature of the experiments allowed the detailed investigation of carefully selected clumps of nanodiamonds over a broad range of ion fluences without the need to remove the support grid from the TEM which might have resulted in alterations to the clump.

*Molecular dynamics*

Molecular dynamics calculations of the effects of heavy ion irradiation of nanodiamonds are performed using independent approaches for the C-C interaction: Environmental Dependent Interaction Potential (EDIP) [37], and the Tersoff-Ziegler-Biersack-Littmark (ZBL) potential [19, 38] as implemented in the open-source LAMMPS Molecular Dynamics simulator [39]. For the Xe-C interaction the two approaches use a Lennard-Jones type potential [40], splined with ZBL for short inter-atomic distances. Both sets of simulation produced qualitatively similar results and agree quantitatively to order of magnitude (for detailed discussion of the numerical differences see below) and are presented together (Figure S1).

*EDIP approach.*

The carbon EDIP methodology, in combination with the Ziegler-Biersack-Littmark (ZBL) potential to describe close approaches, accurately models the behaviour of disordered and amorphous carbons [19]. A nanodiamond grain was modelled as a truncated octahedron [21] with a 2x1 reconstruction of (100) faces to reduce surface energy; initial grain temperature was set at 300 K. The Lennard-Jones potential was employed to model the Xe-C interaction. Due to the significant number of $sp^2$ bonded atoms on the surface, roughly 80% of the atoms in the ND are $sp^3$ bonded; the exact fraction of $sp^3$ atoms varies with particle radius. Incident Xe atoms with energies between 0.7 and 40 keV were directed towards the centre-of-mass of the ND. 25 different directions of incident ions uniformly distributed on the unit sphere were used. Smooth behaviour of the calculated parameters as a function of Xe energy implies that the results are statistically sound. Ejected atoms were excluded from the simulations after 1 ps and the remainder was self-annealed with the residual temperature. To assess thermal contact between particles, a smaller subset of calculations was performed in which the ND undergoing Xe bombardment was itself in contact with another ND of the same size. In typical simulations, around 30 ps elapsed before the two ND's were in thermal equilibrium with each other; the temperature change showed an exponential variation with a time constant of 9 ps. This observation gives an approximate time scale for self-annealing which occurs after Xe impact.

*LAMMPS approach*

In LAMMPS simulations the carbon-carbon interaction was modelled with the Tersoff potential, since it is easy to combine with ZBL to handle large energy interactions. For the Xe-C interactions, a Lennard-Jones (LJ) potential with ε = 121 K and σ = 3.6 Å was used; these parameters were obtained by fitting the dimer and this approach has previously been used for Xe impacts on C nanotubes [40]. Due to the large energies of the Xe atoms, the LJ potential was smoothly "joined" to ZBL at short distances. The nanodiamond grains were created by means of a spherical cut of bulk diamond. The structure was relaxed by energy minimization, followed by annealing at 1800 K during 0.2 ns by velocity rescaling. Finally, the temperature was reduced to 300 K. The Xe impact on 3, 5 and 7 nm nanodiamonds, was studied for incident energies from 4 to 15 keV. An adaptive time step was used, with a minimum of 0.001 fs. For each energy and grain size, 20 collisions were performed in order to develop statistics. Each collision was followed during 10 ps to understand the structure evolution. For very high temperatures (see below) a modified REBO-Scr [41] was used instead of the Tersoff potential, since this takes into consideration the weakening of the C-C interaction when the two C atoms are far apart, in the presence of a third atom. It is relevant to mention that the Tersoff potential is well known to fail when estimating the $sp^2/sp^3$ ratio.

Inspection of Fig. S1 shows that EDIP and the Tersoff potential agree qualitatively on the fact that the highest temperatures are reached with 6 keV energies. However, quantitatively EDIP yields slightly lower values (by approximately 1000 K) than the Tersoff potential. This difference is due to two main factors: i) the ND used in the MD with Tersoff are perfectly spherical, while for EDIP faceted ND were employed; ii) for Tersoff central collisions were simulated, changing slightly the ND initial conditions, to obtain statistics. Therefore, Xe travels a larger distance inside the ND. For EDIP 25 collisions were implemented, with the ND oriented in different directions. While the procedural changes are minor they can yield significant final ND temperature variations, since the Xe ion can deliver different amounts of energy to the ND. In fact, inspection of Fig. 8 shows that changing the impact parameter the final temperature exhibits variations of the order of 1000 K for 10 keV energies. Consequently, in spite of the different methodologies both Tersoff and EDIP are able to capture the basics of the phenomenon, and yield qualitatively consistent results.

*Quantum chemistry*

Behaviour of implanted Xe was investigated in H-terminated non-spherical nanodiamond particles with diameters 1.6, 2.0 and 3.0 nm (326, 649 and 2476 carbon atoms respectively) using quantum chemistry Quickstep module of the CP2K program suite [42] with a dual basis of localized Gaussians and plane waves. The plane wave cutoff was 400 Ry, appropriate for employed Goedecker–Teter–Hutter pseudopotentials [43]. The localized basis set of double zeta plus polarization (DZVP) was quality optimized to reduce the basis set superposition errors [44]. The calculations were performed using the Perdew–Burke–Ernzerhof (PBE) exchange correlation functional [45]. A conjugation gradient (CG) geometry optimization with SCF convergence criteria of $5.0 \times 10^{-7}$ a.u. was used. Atomic configurations were considered converged when forces were less than $4.5 \times 10^{-4}$ hartree×bohr$^{-1}$. The simulations were performed with non-boundary conditions in cubic unit cells which provided distances between diamond particles more than 10 Å to avoid interparticle interaction. Nanoparticles were constructed as described in [34]. After optimization of a pure diamond nanoparticle a grain with a Xe defect was optimized. The Xe defects were built by replacing two neighbouring carbon atoms. Following [34] we calculated the formation energy $E_{form}$ of the defect in nanodiamond using equation 1:

$$E_{form}(q) = E_{tot}(q) - E_{tot}^{"perfect"} + 2\mu_C - \mu_{Xe} \qquad (Eq.\ 1)$$

where $E_{tot}(q)$ – total energy of nanoparticle, $E_{tot}^{"perfect"}$ – total energy of a nanoparticle without defects, $\mu_C(\mu_{Xe})$ – chemical potential of carbon (xenon), can be taken from the calculation for bulk diamond (isolated atom which models ideal gas).

*Acknowledgements*

We highly appreciate provision of Orgueil meteoritic nanodiamonds by A.V. Fisenko and L.F. Semjonova. This work was partly supported by Russian Science Foundation (grant 14-13-01279 to A.A.S.), by FONDECYT project 1160639 (M.K.), CEDENNA (BASAL/CONICYT GRANT FB0807 to M.K.). F.J.V. was supported by CONICYT Doctoral Fellowship 21140948. M.K. was supported by AFOSR Grant FA9550-16-1-0122, and E.M.B. by PICT-2014-0696 (ANPCyT), and a SeCTyP-UNCuyo grant. The MIAMI facility was constructed with funding from EPSRC (grant EP/E017266/1 to S.E.D. and J.A.H.). The quantum chemistry calculations were carried out using high-performance computing resources of the federal collective usage centre Complex for Simulation and Data Processing for Mega-science Facilities at NRC "Kurchatov Institute", http://ckp.nrcki.ru/. This study was supported by the Competitiveness Program of NRNU MEPhI (I.I.V.).

**Author contributions.** A.A.S. designed the study and participated in samples characterisation and TEM investigation. J.A.H., G.G., S.E.D. made the TEM study. N.A.M. and J.F. made the EDIP calculations; F.V., R.G., M.K. and E.B. provided LAMMPS results. A.L.T. performed Xe-V stability calculations. A.A.S. and I.I.V. characterized the samples. All authors took part in discussion of the results and manuscript preparation.

**Competing Financial Interests Statement.**

The authors declare no competing financial interests.